\documentclass[preprint,11pt]{elsarticle}

\usepackage{amssymb}
\newtheorem{theorem}{Theorem}[section]
\newtheorem{lemma}[theorem]{Lemma}
\newtheorem{proposition}[theorem]{Proposition}
\newtheorem{corollary}[theorem]{Corollary}

\newenvironment{definition}[1][Definition]{\begin{trivlist}
\item[\hskip \labelsep {\bfseries #1}]}{\end{trivlist}}
\newenvironment{example}[1][Example]{\begin{trivlist}
\item[\hskip \labelsep {\bfseries #1}]}{\end{trivlist}}

\newcommand{\R}{{\cal R}}
\newcommand{\Z}{{\cal Z}_0}
\newcommand{\M}{{\cal M}}
\newcommand{\N}{{\cal N}}
\newcommand{\bM}{{\bf M}}

\journal{Applied Mathematics and Computation}

\begin{document}

\begin{frontmatter}


%%\title{Title\tnoteref{label1}}

\title{On asymptotic solutions of Friedmann equations}


\author[label1]{\v Z. Mijajlovi\'c\corref{corr1}}
\ead{zarkom@matf.bg.ac.rs}
\cortext[corr1]{Corresponding author}
\author[label1]{N. Pejovi\'c}
\author[label1]{S. \v{S}egan}
\author[label2]{G. Damljanovi\'c}

\address[label1]{Faculty of Mathematics, Univ. of Belgrade, Studentski trg 16,  Serbia}
\address[label2]{ Astronomical Observatory in Belgrade,  Volgina 7,   Serbia}

\begin{abstract}
   Our main aim  is to apply the theory of regularly varying functions
   to the asymptotical analysis at infinity of solutions of
   Friedmann cosmological equations.
   A new constant $\Gamma$ is introduced  related to the Friedmann cosmological  equations.
   Determining the values of   $\Gamma$ we obtain the asymptotical behavior of the
   solutions, i.e.  of the expansion scale factor $a(t)$ of a universe.
   The instance $\Gamma<\frac{1}{4}$ is appropriate for both cases,
   the spatially flat and open universe, and
   gives a sufficient and necessary condition for the solutions to
   be regularly varying. This property of Friedmann equations is formulated as the
   {$generalized\ power\ law\ principle.\ $}
   From the theory of regular variation
   it follows that the solutions under usual assumptions include a multiplicative term which
   is a slowly varying function.
   %Under usual assumptions  for the scale factor $a(t)$, it turns out that this slowly varying term exists.
\end{abstract}


\end{frontmatter}

% \linenumbers

\section{Introduction}
\label{}

 In this paper\footnote{This work is partially supported by the Serbian Ministry of Science, Grant III 44006.}
 we describe conditions under which the Friedmann
 equations \cite{5} have regularly varying solutions.
 Strictly speaking, we found a necessary and sufficient condition for Friedmann equations, expressed by the values of a constant $\Gamma$,
 to have regularly varying solutions.
 We formulate this description  as the {\sl generalized power law principle for Friedmanns equation}.
 The physical formulation of this  condition is
 that a certain form of the equation of state $p\sim w\rho c^2$ must hold.
 Hence, our discussion is mainly about a universe filled with the perfect fluid with constant barotropic
 equation of state $p= w\rho c^2$.
 The sufficiency of this condition is well known, e.g.  Liddle and Lyth \cite{9}, Coles and Lucchin \cite{4}, Narlikar \cite{15}
 and Islam \cite{8}.
 However, we have not found in the literature the necessity part of the power law principle.

 It appears that the mentioned constant $\Gamma$
 related to the Friedmann acceleration equation  plays the crucial role in this analysis.
 Its values determine the asymptotical behavior  of the
 solutions of the Friedmann equations, i.e.  of the  scale factor $a(t)$
 as time $t$ tends to $\infty$. Our solution is also valid for non-zero cosmological constant $\Lambda$
 if the pressureless spatially flat universe is assumed. This was possible due to a formula of
 Carroll at al. \cite{3} for the predicted age of the universe.
 In the course of this analysis, mathematical singularities appearing in the solutions
 are classified and are clearly distinguished from those arising from the
 physical constrains.
 All solutions we found are in agreement with the results widely found in the literature on standard cosmological model.

 The background for our analysis is the theory of regularly varying functions which could be considered as the
 mathematical counterpart of the general form of the power law, the term often used in physics. A good presentation of this subject can be
 found in Bingham at al. \cite{2} and Seneta \cite{17}. Another tool we used is the theory of regularly varying solutions of differential
 equations. A good source for this theory is Mari\'c \cite{12}.
 The theory of regular variation
 provides additional means in the asymptotical analysis of the solutions of the second
 order linear differential equations as  (\ref{AQ}), but it seems it has not been much applied  in cosmology
 and in astrophysics. There are few such applications,  e.g. Molchanov at al. \cite{14},
 Stern \cite{18} and Mijajlovic at al. \cite{13}.

 By $\mathbb{R}$ we denote the set of real numbers. As usually, for two real functions $f$ and $g$, $f(x)\sim g(x)$
 (or $f \sim g$)  means that $\lim_{x\to\infty} f(x)/g(x)=1$.

 The paper is organized as follows. In the first section   the history of the problem is explained and
 physical (Friedmann equations) and mathematical (regular variation) background is given. The main results of the paper
 are presented in Sections 2 and 3.

%__________________________________________________________________

 \subsection{Friedmann equations}

 The   scale factor $a(t)$ is defined by Friedmann-Lema\^{i}tre-Robertson-Walker (FLRW) metric.
 The FLRW 4-dimensional line element in
 spherical comoving coordinates is given by
 \begin{equation}\label{FLRW}
   ds^2= -dt^2 + a^2(t)\left[\frac{dr^2}{1-kr^2} + r^2(d{\theta}^2 + \sin^2{\theta}d\varphi^2)\right]
 \end{equation}
 This metric is an exact solution of Einstein's field equations of general relativity and
 it describes a homogeneous, isotropic expanding or contracting universe.
 In this paper we shall discuss only the expanding universe.
 The scale factor $a(t)$ is a solution of the Friedmann equations. These equations are derived from the Einstein field
 equations; they are the following three differential equations.
 The term Friedmann equation is usually reserved for the first one.
\begin{equation}\label{Friedman}
   \left(\frac{\dot{a}}{a}\right)^2= \frac{8\pi G}{3}\rho  -\frac{kc^2}{a^2}
 \end{equation}
 The Friedmann acceleration equation is
 \begin{equation}\label{AQ}
   \frac{\ddot{a}}{a}= -\frac{4\pi G}{3}\left(\rho + \frac{3p}{c^2}\right)
 \end{equation}
 while the fluid equation is
 \begin{equation}\label{fluid}
   \dot\rho + 3\frac{\dot{a}}{a} \left(\rho + \frac{p}{c^2}\right)= 0.
 \end{equation}
 The solutions of these equations are three fundamental parameters,
 the scale factor $a=a(t)$, the energy density $\rho=\rho(t)$
 and $p=p(t)$, the pressure of the material in the universe.
 Here $k$ is the curvature index with possible values $1$ (elliptic geometry), 0 (spatially flat geometry) and $-1$ (hyperbolic geometry).
 The symbol $G$ denotes the gravitational constant and $c$ is the speed of light.
 Equations (\ref{Friedman}) -- (\ref{fluid}) are not independent.
 Eq. (\ref{AQ}) follows from (\ref{Friedman}) and (\ref{fluid}),
 while the Eqs. (\ref{Friedman}) and (\ref{AQ}) yield (\ref{fluid}).

 We shall use Karamata theory of regularly varying functions, as applied to differential equations in
 Mari\'c and Tomi\'c [12] and Mari\'c \cite{12}.
 This theory generalizes the power law in physics and we shall use it to obtain the asymptotic analysis
 of solutions of Friedmann equations.

 In our study of the asymptotical solutions of Friedmann equations, the acceleration equation will have the
 central point for several reasons. First, it does not contain explicitly the curvature index $k$. Secondly,
 the theory of regularly varying solutions of such type of equations
 can be applied successfully, regardless if the cosmological constant $\Lambda$ is added in (\ref{Friedman}) and (\ref{AQ}):
 \begin{equation}\label{AQL}
   \left(\frac{\dot{a}}{a}\right)^2= \frac{8\pi G}{3}\rho  -\frac{kc^2}{a^2} + \frac{\Lambda}{3},\hskip 6mm
   \frac{\ddot{a}}{a}= -\frac{4\pi G}{3}\left(\rho + \frac{3p}{c^2}\right) + \frac{\Lambda}{3}.
 \end{equation}

 Namely, under the transformations
  $ \rho'= \rho + \Lambda/(8\pi G), \enskip
     p'  =  p   - \Lambda/(8\pi G) $
 the equations  (\ref{AQL}) yield (\ref{Friedman}) and (\ref{AQ}), but now with respect to the parameters
 $\rho'$ and $p'$. The fluid equation is not affected by the parameter $\Lambda$.
 Therefore, our discussion will be concentrated further on the solutions of the Friedmann equations in their basic form
  (\ref{Friedman}) -- (\ref{fluid}) if it is not otherwise stated.

 From now on, we shall assume that the functions $a(t)$, $p(t)$ and $\rho(t)$  satisfy all three Friedmann equations.
 %(\ref{Friedman}), (\ref{AQ}) and (\ref{fluid}).
 We shall also assume that all appearing functions are continuous in their domains and have the sufficient number of
 derivatives, at least that they have the continuous second derivative.

% Our approach to the analysis of the Friedmann equations is as follows.
% Assume $\bar p$ and $\bar \rho$ are solutions of Friedmann equations and let $A(a,\bar p,\bar \rho)=0$ be
% the acceleration equation (\ref{AQ}) where the unknown quantities $p$ and $\rho$ are replaced by
% $\bar p$ and $\bar \rho$. Then every solution $\bar a(t)$ of Friedmann equations is, of course, a
% solution of the acceleration equation. Hence all conclusions obtained for the general solution $a(t)$
% of $A(a,\bar p,\bar \rho)=0$ are also valid for the solutions $\bar a(t)$ of all three Friedmann equations.
% For example, if the general solution of $A(a,\bar p,\bar \rho)=0$ is regularly varying,
% then the solutions of Friedmann equations are regularly varying, too. Hence, from now on we shall
% assume that $p$ and $\rho$ satisfy Friedmann equations.

 % The theory of regularly varying functions
 % provides additional means in the asymptotical analysis of the solutions of the second
 % order linear differential equations as  (\ref{AQ}). This theory is quite well developed in Seneta \cite{17},
 % Bingham \cite{2} and Mari\'c \cite{12}, but it seems it has not been much applied  in cosmology
 % and in astrophysics in general. However, there are few such applications,  e.g. Molchanov \cite{14},
 % Stern \cite{18} and Mijajlovic \cite{13}.

 %__________________________________________________________________

  \subsection{Regular variation}

 In this section we shall  review the basic notions related to the
 regular variation necessary for our analysis.  In particular we shall need properties of regularly varying
 solutions of the second order differential equation %of the form
 \begin{equation}\label{SE}
           \ddot{y} + f(t)y= 0,\quad f(t)\,\, {\rm is\,\, continuous\,\, on}\,\, [\alpha,\infty].
 \end{equation}
 Observe that the acceleration equation (\ref{AQ}) has the form (\ref{SE}).
 In short, the notion of a regular variation is related to the power law distributions, described by the following
 relationship between quantities $F$ and $t$: %It exhibits the property of scale invariance represented by
 \begin{equation}
           F(t)= t^r(\alpha + o(1)), \quad \alpha, r \in \mathbb{R}. %\hskip 2mm {\rm are \hskip 1mm real\hskip 1mm constants.}
 \end{equation}
 %The term $o(t^k)$ is usually neglected in formulations of physical laws.
 %The constant $r$ is called the scaling exponent.

%The use of the notion of power law  in the literature is somewhat ambiguous.
%The most simple form of the power law is given by the equation $y=t^k$.
 It is said that two quantities $y$ and $t^r$ satisfy the power law if they are related by
 a proportion,\footnote{This relation is usually denoted by $y\propto t^r$.}
 i.e. there is a constant $\alpha$ so that $y= \alpha t^r$.
 This definition of power law can be naturally extended by use of the notion of
 slowly varying function.  %introduced by J. Karamata \cite{10}.

  A real positive continuous
  function\footnote{More generally it may be assumed that $L(t)$ is a measurable function,
  but in this article we are dealing only with continuous functions anyway.}
  $L(t)$ defined for $x>x_0$  which satisfies
  \begin{equation}\label{KI}
    \frac{L(\lambda t)}{L(t)}\to 1\quad  {\rm as}\quad t\to \infty, \quad
          \textstyle{\rm for \hskip 1mm  each \hskip 1mm real}\hskip 1mm \lambda>0.
 \end{equation}
 is called a slowly varying function.

 \begin{definition}
 A physical quantity F(t) is said to satisfy the  generalized power law if
 \begin{equation}\label{GPL}
    F(t)= t^rL(t)
 \end{equation}
  where $L(t)$ is a slowly varying function and $r$ is a real constant.
 \end{definition}

 Examples of slowly varying functions are
 $\ln(x)$ and iterated logarithmic functions $\ln(\ldots\ln(x)\ldots)$.
 More complicated examples (cf. Mari\'c \cite{12})  are provided by:
 \begin{equation}\label{erdos}
   L_1(x)= \frac{1}{x}\int_a^x \frac{dt}{\ln t}, \quad L_2(x)= {\rm exp}((\ln x)^{1/3}\cos(\ln x)^{1/3})
 \end{equation}
 We note that $L_2(x)$ varies infinitely between 0 and $\infty$.

 A positive continuous function $F$ defined for $t>t_0$, %where $a$ is a constant,
 is the regularly varying function of the index $r$,  if and only if it satisfies
 \begin{equation}\label{KII}
    \frac{F(\lambda t)}{F(t)}\to {\lambda}^r \quad {\rm as}\quad t\to \infty,\quad
          \textstyle{\rm for \hskip 1mm  each}\hskip 1mm \lambda>0.
 \end{equation}
 It immediately follows that a regularly varying function $F(t)$ has the
 form (\ref{GPL}).
 %\begin{equation}
 %    F(t)= L(t)t^r
 %\end{equation}
 %where $L(t)$ is slowly regular.
 So to say that $F(t)$
 is regularly varying is the same as $F(t)$ to satisfy the generalized power law.
 %In this article, whenever we refer to power law we mean the generalized power law.
 By Proposition 7 in \cite{12}, if a function $F(x)$ is asymptotically equivalent to a regularly varying function, {\it it is}
 a regularly varying function. Hence, we may define the generalized power law also by
 \begin{equation}
   F(x)\sim t^\alpha L(t),\quad {\rm as}\,\, t\to\infty.
 \end{equation}

 The class of regularly varying functions of index $\alpha$
 we shall denote by $\cal R_\alpha$. Hence ${\cal R}_0$ is the class of all slowly varying functions.
 By ${\cal Z}_0$ we shall denote the class of zero functions at $\infty$, i.e.
 $\varepsilon\in {\cal Z}_0$ if and only if $\displaystyle\lim_{t\to +\infty} \varepsilon(t)=0$.

 J. Karamata  introduced in \cite{10} the concept of regularly
 varying functions continuing the works of G.H. Hardy, J.L. Littlewood and  E. Landau in the
 asymptotic analysis of real functions.
 The following two theorems describe fundamental properties of this class of functions.
 \vskip 2mm
  \begin{theorem}\label{KFT1} {\rm\cite{10}} {\rm(}Representation theorem{\rm )}
  $L\in\R_0$  if and only if there are measurable
  functions $h(x)$ and $\varepsilon\in\Z$ and $b\in \mathbb{R}$ so that
  \begin{equation}\label{RVrepresentation}
     L(x)= h(x) e^{\int_b^x\frac{\varepsilon(t)}{t}dt}, \quad x\geq b,
  \end{equation}
  and $ h(x)\to h_0$ as $x\to\infty$, $h_0$ is a positive constant.
  \end{theorem}

 %Slowly varying functions have the following representation (Karamata \cite{10}):
 %\begin{equation}\label{RVrepresentation}
 %    L(x)= h(x) e^{\int_b^x\frac{\epsilon(t)}{t}dt}, \quad x\geq b,
 %\end{equation}
 %where $\epsilon(x)\to 0$ and $ h(x)\to h_0$ as $x\to\infty$,
 %$h_0$ is a positive constant.
 The function $\varepsilon(t)$
 in the above theorem is not uniquely determined.
 %In fact $\varepsilon(t)$ can be chosen so that it has the $k$-th derivative up to the arbitrary positive integer $n$.
 If $h(x)$ is a constant function,  then $L(x)$ is called normalized. We denote by $\cal N$
 the class of normalized slowly varying functions.
 We note the following important fact for $\cal N$-functions. If $L\in \cal N$ and there is $\ddot L$,
 then $\varepsilon$ in (\ref{RVrepresentation}) has the first order derivative $\dot\varepsilon$.
 This follows from the identity $\varepsilon(t)= t\dot L(t)/L(t)$.

 % In particular,
 % we can take for an appropriate constant $\lambda_0$ (cf. Seneta \cite{17} and Bingham \cite{2}):
 % \begin{equation}\label{epsilon}
 %    \epsilon(x) = \frac{1}{\log(\lambda_0)}\log\left(\frac{L(\lambda_0x)}{L(x)}\right).
 % \end{equation}

 % \begin{theorem}\label{KFT2} {\rm\cite{10}} {\rm(}Uniform convergence theorem{\rm )}
 %  If $F$ is regularly varying of index $\alpha\in \mathbb{R}$ then the relation {\rm(\ref{KII})}
 % holds uniformly for $\lambda\in [b_1,b_2]$ with $0<b_1<b_2<\infty$. If $\alpha< 0$, then
 % the uniform convergence holds  for $b_1\leq\lambda<\infty$.
 % \end{theorem}

  There is also an appropriate definition of regular variation at 0 and $\infty$ and various generalizations
  such as the rapidly varying functions. Even if such solutions of Friedmann equation are possible,
  we will not discuss these types of solutions in this article, so we omit these definitions.

  % One can find the full theory of these functions and related notions in
  % Seneta \cite{17} and Bingham \cite{2}.

  For our study of Friedmann equations we need several results on solutions of (\ref{SE}).
  % $\ddot{y} + f(t)y=0$, $f(t)$ is continuous on $[\alpha,\infty]$.
  %% The reason is that the acceleration equation (\ref{AQ})  is exactly of this type.
  There are various conditions for  $f(t)$ that ensure that regularly varying solutions
  of  $\ddot{y} + f(t)y=0$ exist.
  % The very good treatise of this subject is given in Mari{\' c} \cite{12}.
  We shall particularly use  the following result, see  Howard and Mari\'c [7] and Mari\'c \cite{12} the theorems 1.10 and 1.11:
  \vskip 2mm
  \begin{theorem}\label{MaricT}

  $Let\ -\infty< \Gamma <1/4,\  and\ let\ \alpha_1<\alpha_2\ be\ two\
  roots\  o\!f  the\ equa\-tion$
  \begin{equation}\label{EQ144}
      x^2 - x +\Gamma= 0.
  \end{equation}
  Further let $L_i$, i=1,2 denote two normalized slowly varying functions.
  Then there are two linearly independent regularly varying solutions of\, $\ddot{y} + f(t)y=0$\, of the
  form
  \begin{equation}\label{EQ145}
         y_i(t) = t^{\alpha_i}L_i(t), \quad i=1,2,
  \end{equation}
  if and only if $\displaystyle \lim_{x\to\infty} x\int_x^{\infty}\hskip -2mm f(t)dt= \Gamma$.
  Moreover,  $\displaystyle L_2(t)\sim \frac{1}{(1-2\alpha_1)L_1(t)}$. \qed

  \end{theorem}
  \vskip 2mm

  The limit of the integral in the theorem is not always easy to compute.
% The following Corollary leads to more applicable form of the previous theorem.
%%  If we apply L'Hospital rule to the limit
%%  %\smallskip
%%
%%  $$\displaystyle \lim_{x\to\infty} x\int_x^{\infty}\hskip -2mm f(t)dt= \lim_{x\to\infty} \int_x^{\infty}\hskip -2mm f(t)dt/(1/x).$$
%%  %\smallskip
%%  \noindent
%%  then $ \displaystyle \lim_{t\to\infty}t^2 f(t)= \Gamma$ implies
%%   $\displaystyle \lim_{x\to\infty} x\int_x^{\infty}\hskip -2mm f(t)dt= \Gamma$.
%%  %\vskip 2mm
%%  Combining this implication with Theorem 1, we obtain a somewhat weaker, but useful form of Theorem 1:
%   Namely, by simple use of L'Hospital rule we see that
%   $ \displaystyle \lim_{t\to\infty}t^2 f(t)= \Gamma$ implies
%   $\displaystyle \lim_{x\to\infty} x\int_x^{\infty}\hskip -2mm f(t)dt= \Gamma$.
%   Therefore, we obtain a somewhat weaker, but useful form of Theorem 3.1:
%  \vskip 2mm
%
%  \begin{corollary} {Let $-\infty< \Gamma <1/4$ and let $\alpha_1<\alpha_2$ be two roots
%  of the equation
%  \begin{equation}\label{EQ146}
%      x^2 - x +\Gamma= 0.
%  \end{equation}
%  Further let\quad $L_i$, i=1,2, denote two normalized slowly varying functions.
%  If\quad $\displaystyle \lim_{t\to\infty} t^2f(t)= \Gamma$
%  then there exist two linearly independent regularly varying solutions of (\ref{AQ}) of the
%  form
%  \begin{equation}\label{EQ147}
%         y_i(t) = t^{\alpha_i}L_i(t), \quad i=1,2.
%  \end{equation}
%   Moreover, one has $\displaystyle L_2(t)\sim \frac{1}{(1-2\alpha_1)L_1(t)}$.} \hfill
%   \qed
%
%   \end{corollary}
   As $\displaystyle \lim_{t\to\infty}t^2 f(t)= \Gamma$ implies
   $\displaystyle \lim_{x\to\infty} x\int_x^{\infty}\hskip -2mm f(t)dt= \Gamma$,
   we see that
   \begin{equation}\label{WeakC}
    \lim_{t\to\infty}t^2 f(t)= \Gamma
   \end{equation}
   gives a  useful sufficient condition for the existence of regular solutions of the equation
   $\ddot{y} + f(t)y=0$ as described in the previous theorem.
   \vskip 2mm

 %__________________________________________________________________

 \section{Regularly varying solutions of acceleration equations}

  %%\section{Power solutions of acceleration \\
  %%         \hskip 6mm equation}

  As noted, the acceleration equation obviously has the form (\ref{SE}) so under
  appropriate assumptions, i.e. that the functions we encounter are continuously differentiable as many times  as necessary,
  the analysis of the previous section, in particular the theorem \ref{MaricT},
  can be applied to it. For this reason, we shall write from now on
  the acceleration equation (\ref{AQL})  in the form
  \begin{equation}\label{AQL2}
     \ddot a + \frac{\mu(t)}{t^2}a=0,
  \end{equation}
  where
  \begin{equation}\label{mu}
    %\mu(t)= t^2\left(\frac{4\pi G}{3}\left(\rho + \frac{3p}{c^2}\right) - \frac{\Lambda}{3}\right).
    \mu(t)=  \frac{4\pi G}{3}t^2\left(\rho + \frac{3p}{c^2}\right).
  \end{equation}
  Our approach in the next analysis is as follows.
  Obviously $\mu$ is a function of $\rho$ and $p$. We assumed that $\rho$ and $p$ are  solutions of
  Friedmann equations, hence $\mu(t)$ is a well-defined function. Under this assumption, the theory of regular variation
  applied to the equation (\ref{AQL2}) yields the asymptotic expansions of
  $\mu(t)$ and $a(t)$ and exact conditions on $\mu(t)$ under which these expansions exist. Using the identity
  (\ref{mu}) %and that $a(t)$, $\rho(t)$ and $p(t)$ satisfy Friedmann equations,
   we will be able then to find the asymptotical expansions for $\rho$, $p$ and other
  cosmological parameters.

  %By  Theorem \ref{MaricT} and the note after it, if $f(t)= \mu(t)/t^2$ we have immediately:
  %\vskip 2mm

  In the next the crucial role will play the following integral limit:

  %\begin{corollary}\label{MaricT2}
  %Assuming   conditions of Theorem {\rm \ref{MaricT}},
  %%and\ let\ \alpha_1<\alpha_2\ be\ two\  roots\  of\ the\ equa\-tion$
  %%\begin{equation}\label{EQ144}
  %%   x^2 - x +\Gamma= 0.
  %%\end{equation}
  %there are solutions of the equation {\rm (\ref{AQL2})} that obey the generalized power law
  %%Further let $L_i$, i=1,2 denote two normalized slowly varying functions.
  %%Then there are two linearly independent regularly varying solutions of\, $\ddot{y} + f(t)y=0$\, of the
  %%form
  %%\begin{equation}\label{EQ145}
  %%       y_i(t) = t^{\alpha_i}L_i(t), \quad i=1,2,
  %%\end{equation}
  %if and only if
  \begin{equation}\label{gamma}
  \displaystyle \lim_{x\to\infty} x\int_x^{\infty}\hskip -1mm \frac{\mu(t)}{t^2}dt= \Gamma.
  \end{equation}
  %%Moreover,  $\displaystyle L_2(t)\sim \frac{1}{(1-2\alpha_1)L_1(t)}$.\qed

  %The existence of the limit $\displaystyle \lim_{t\to\infty}\mu(t)= \Gamma$ is a sufficient condition for
  %the existence of the solutions of
  %{\rm (\ref{AQL2})} that satisfy the generalized power law. \qed

  %\end{corollary}
  %\vskip 2mm

  Let us denote by $\M_\Gamma$ the class of functions $\mu$ that satisfy the integral condition (\ref{gamma}).
  Further, let
  $$\M = \bigcup_{r\in\mathbb{R}}\M_r.$$
  Mari\'c introduced the integral condition (\ref{gamma}) (cf. \cite{12}), accordingly we shall  call the class $\M$ also
  as Mari\'c class of functions.
  Obviously, $\M$ is a vector space over $\mathbb{R}$ and the map $\bM: \M \rightarrow \mathbb{R}$
  defined by
  $$
     \bM(u)= \lim_{x\to\infty} x\int_x^{\infty}\hskip -1mm \frac{u(t)}{t^2}dt
  $$
  \noindent
  is a linear functional, i.e. $\bM(\alpha u + \beta v)  = \alpha\bM(u) + \beta \bM(v)$, $\alpha,\beta\in \mathbb{R}$, $u,v\in \M$.
  It is easy to see that $\bM(\varepsilon) = 0$ for $\varepsilon\in\Z$.
  By the note regarding (\ref{WeakC}), we immediately  have

  \begin{proposition}\label{PropositionM}
     \quad If\, $\displaystyle\lim_{t\to \infty}u(t)=r$   then $\bM(u)= r$.
  \end{proposition}

  % The converse of this proposition in certain cases is also true.

  %% $\noindent{\bf Proof}\, Assume $\varepsilon\in \Z$. Then $\int_x^\infty \varepsilon(t)/t^2dt= o(1/x)$, hence
  %% $\bM(\varepsilon)=0$. If $\displaystyle\lim_{t\to \infty}u(t)=r$ then $u(t) = r + \varepsilon(t)$ for some
  %% $\varepsilon\in \Z$, so

  %% $\bM(u)= \bM(r+\varepsilon)= \bM(r) + \bM(\varepsilon) =r$. \hfill$\bigtriangleup$
  %% \vskip 2mm

  % \begin{theorem}
  %  Suppose $u(t)$ is regularly varying and  $\bM(u)=r$ for some $r\in \mathbb{R}$.
  %  Then $\displaystyle \lim_{t\to\infty} u(t)=r$.
  % \end{theorem}

  % \noindent{\bf Proof}\quad We give proof only for slowly varying functions. So suppose
  % $u\in \R_0$ and $\bM(u)=r$. Let $0<\delta <1$. Then $v(t)= u(t)t^{-\delta}$  varies regularly with index $-\delta<0$ and
  % $$
  %   I(x)\equiv x\int_x^\infty \frac{u(t)}{t^2}dt =
  %   u(x)\left( \int_1^\infty\frac{1}{s^{2-\delta}}\left(\frac{v(sx)}{v(x)} - s^{-\delta}\right)ds + \int_1^\infty \frac{ds}{s^2}\right).
  % $$
  % By Theorem \ref{KFT2}, $\displaystyle\frac{v(sx)}{v(x)} - s^{-\delta} \to 0$ uniformly as $x\to \infty$ and
  % $\displaystyle g(s)=\frac{1}{s^{2-\delta}}$ is integrable
  % over $[1,\infty)$. Hence the first integral on the right side of above relation  converges to $0$, i.e.
  % $$
  %   I(x)= u(x)(\varepsilon(x) +1), \quad \varepsilon\in \Z.
  % $$
  % From the last relation immediately follows $\displaystyle \lim_{x\to\infty} u(x)= \lim_{x\to\infty} I(x)= r$. \hfill $\square$
  % \vskip 2mm

  Now we prove a useful representation theorem for Mari\'c class of functions.

  \begin{theorem}\label{RTM} {\rm (Representation theorem for $\cal M$-functions)} $u\in \M_r$ if and only if there are
    $\varepsilon, \eta \in \Z$ such that $u(t)= r - t\dot\varepsilon(t) +\eta(t)$. If $r<1/4$ then $\varepsilon$ is
    that one appearing in the representation {\rm (\ref{RVrepresentation})} of $a(t)$, with $h(t)$  constant. %Theorem {\rm \ref{MaricT}}.
  \end{theorem}

  \noindent{\bf Proof}\, ($\Rightarrow$)\, Suppose $u\in\M_r$, and $\displaystyle r< \frac{1}{4}$.
  By Theorem \ref{MaricT} the equation $\displaystyle \ddot y + \frac{u(t)}{t^2}y = 0$ has a
  solution $a(t)= t^\alpha L(t)$ where $\alpha$ is a root of the equation $x^2 -x +r= 0$ and $L\in\N$.
  By Theorem \ref{KFT1} there are $a_0,b\in \mathbb{R}$ and $\varepsilon\in\Z$ so that
  $ a(t)= a_0t^\alpha e^{\int_b^x\frac{\varepsilon(t)}{t}dt}$. As $\displaystyle\dot L(t) = \frac{\varepsilon(t)}{t}L(t)$
  and $r= -\alpha(\alpha - 1)$, we have
  $$
     \ddot a(t) = (-r + t\dot\varepsilon - (1-2\alpha)\varepsilon +\varepsilon^2)L(t)t^{\alpha -2}
  $$
  Since
  $\displaystyle
     -\frac{u(t)}{t^2}= \frac{\ddot a(t)}{a(t)}
  $
  it follows
  $
     u(t)= r - t\dot\varepsilon(t) +\eta(t)
  $
  where $\eta =  \varepsilon^2 - (1-2\alpha)\varepsilon$.

  Suppose $\displaystyle r\geq \frac{1}{4}$. As $\bM(\frac{1}{8r}u) = \frac{1}{8}$,
  taking $\displaystyle\frac{1}{8r}u$ instead of $u$ in the  previous proof, we have
  $\frac{1}{8r}u(t)=  \frac{1}{8}- t\dot\varepsilon(t) +\eta(t)$ for some $\varepsilon,\eta\in\Z$,
  hence
  \vskip 1mm
  $
     u(t)= r - t\dot\varepsilon_1(t) +\eta_1(t)
  $
  where $\varepsilon_1 = 8r\varepsilon$ and $\eta_1 = 8r\eta$.

  \noindent ($\Leftarrow$)\,
  Suppose
  $
     u(t)= r - t\dot\varepsilon(t) +\eta(t)
  $
  where $\varepsilon, \eta \in \Z$. Then
  $$
    \bM(u) = \bM(r) - \bM(t\dot\varepsilon) + \bM(\eta) = r - \bM(t\dot\varepsilon).
  $$
  Further, taking $v(t)= t\dot\varepsilon$,
  $$
      \int_x^\infty \frac{v(t)}{t^2}dt = \int_x^\infty \frac{d\varepsilon}{t}=
          -\frac{\varepsilon}{x} + \int_x^\infty \frac{\varepsilon}{t^2}dt = -\frac{\varepsilon}{x} + o\left(\frac{1}{x}\right).
  $$
  Hence,
  $\displaystyle\bM(t\dot\varepsilon) = \lim_{x\to\infty} x\int_x^\infty \frac{v(t)}{t^2}dt =
           \lim_{x\to\infty}(-\varepsilon + o(1)) = 0$, so $\bM(u)=r$. \qed
 %\hfill $\square$
  \vskip 2mm

  \begin{corollary}\label{EpsC}
     Assume $u\in\M_r$. Then  $\displaystyle\lim_{t\to\infty} u(t)=r$ if and only if\,

     \noindent
     $\displaystyle\lim_{t\to\infty} t\dot\varepsilon(t) = 0$ in above representation of $u$.
  \end{corollary}

  \begin{example} Let $\displaystyle\varepsilon(t) = \frac{\sin(t^3)}{t}$. Then for
    $$
      \mu(t)= \frac{1}{8} -t\dot\varepsilon(t) - \varepsilon(t)= \frac{1}{8} - 3t^2\cos(t^3)
    $$
    $\displaystyle\bM(\mu) = \frac{1}{8}$ and all ultimately positive solutions of (\ref{AQL2}) are
    regularly varying, but  $\displaystyle\lim_{t\to\infty} \mu(t)$ does not exist.
    Note that it follows $\displaystyle\lim_{x\to\infty} x\int_x^\infty \cos(t^3) dt= 0$.
    \hfill{$\square$}
  \end{example}

  The next proposition will be useful in our further analysis. It also gives the $\varepsilon$-representation
  of the logarithmic derivative $H(t)\equiv \dot a(t)/a(t)$ of $a(t)$.

  \begin{proposition}\label{MaricT3}
    Suppose $\mu\in\M$ and $\displaystyle\Gamma\equiv\bM(\mu) < \frac{1}{4}$  holds for Eq. {\rm (\ref{AQL2})}.
    Then any ultimately positive solution $a(t)$
    of {\rm (\ref{AQL2})} is a normalized regularly varying function, i.e. there are $L\in\N$ and $\alpha\in \mathbb{R}$,
    so that $a(t)= t^\alpha L(t)$.

    If $L(t)$ has the   $\varepsilon$-representation as in Theorem {\rm \ref{KFT1}}, where $h(t)$ is a positive constant,
   then $H(t)= \alpha/t + \varepsilon/t$.
  \end{proposition}

  \noindent{\bf Proof}\quad Suppose $a(t)$ is positive at $\infty$.
  By Theorem \ref{MaricT} there are   $L_1,L_2\in \N$    and
   $\alpha_1,\alpha_2\in \mathbb{R}$  so that
  \begin{equation}\label{GeneralSolution}
     a(t)= c_1L_1t^{\alpha_1} + c_2L_2t^{\alpha_2}, \quad c_1,c_2\in \mathbb{R},
  \end{equation}
  where $\alpha_1,\alpha_2$ are the roots of the equation (\ref{EQ144}). Since $\Gamma<1/4$  we have $\alpha_1\not=
  \alpha_2$, so we may assume $\alpha_1> \alpha_2$. Suppose $c_1\not=0$.
  Hence, the term $c_1L_1t^{\alpha_1}$ dominates $c_2L_2t^{\alpha_2}$, so there is $t_0>0$ so that
  $a(t)> 0$ for $t>t_0$. Let $\delta= \alpha_2-\alpha_1$, $c_0=c_1/c_2$ and $L_0= L_1/L_2$.
  Note that $L_0\in\R_0$.
  By Representation theorem \ref{KFT1} and as $L_1,L_2$ are normalized,
  there are constants $h_1,h_2,b \in \mathbb{R}$ and $\varepsilon_1,\varepsilon_2\in\Z$ so that
  $$
    L_i(x)= h_i e^{\int_b^x\frac{\varepsilon_i(t)}{t}dt}, \quad x\geq b,\quad i=1,2.
  $$
     As $\displaystyle\dot L_i(t)= \frac{\varepsilon_i(t)}{t}L_i(t)$,
     taking the logarithmic derivative  $H(t)\equiv \displaystyle\frac{\dot a(t)}{a(t)}$\, of $a(t)$ and $\alpha\equiv\alpha_1$ we obtain
  $$
     H(t) =
     \frac{\alpha}{t} (1+\varepsilon_1/\alpha)
     \frac{1 + c_0  \displaystyle\frac{\alpha_2+\varepsilon_2}{\alpha_1+\varepsilon_1}L_0(t)t^\delta}
     {1+c_0L_0(t)t^\delta}
  $$
  Since $L_0$ is slowly varying and $\delta<0$, it follows $L_0t^\delta\to 0$ as $t\to 0$. Hence,
  there is $\varepsilon\in\Z$ so that
  \begin{equation}\label{Hubble}
    H(t) = \frac{\dot a(t)}{a(t)}=\frac{\alpha}{t} + \frac{\varepsilon(t)}{t}.
  \end{equation}
  By integration of this relation we have immediately
  $$
        a(x)= a_0e^{\int_{b}^x\frac{\varepsilon(t)}{t}dt}, \quad x\geq b,\quad {\rm where}\,\, b=t_0, a_0= a(t_0).
  $$
  Hence, by Theorem \ref{KFT1},  $a(t)$ is a normalized slowly varying function. \qed
  %\hfill $\square$
  %\newline\phantom{.}   \hfill $\square$
  \vskip 2mm

  %According to the theorem \ref{MaricT} the Friedmann acceleration equation (\ref{AQ}) has two different
  %fundamental solutions  that satisfy  a po\-wer law if and only if the condition (\ref{gamma}) holds.
  %For description of these solutions we shall rely on the theorem \ref{MaricT}.

  %--------------------------------
  %--------------------------------

  \section{Asymptotic solutions of Friedmann equations}

  We proceed to the analysis of
  solutions of Friedman equations taking into account the physical constraints.
  We remind that  besides the acceleration equation (\ref{AQL2}) with $\mu(t)$
  defined by (\ref{mu}), the
  scale factor $a(t)$ also  satisfies the other two Friedmann equations  (\ref{Friedman}) and (\ref{fluid}).
  Note that $a(t)$  is a function of time which represents the relative expansion of the universe.
  This function relates the proper distance between a pair of objects,
  e.g. two galaxies, moving with the Hubble flow in a  FLWR
  universe at any arbitrary time $t$ to their distance at some reference time $t_0$. Thus, $d(t)=a(t)d(t_0)$ where
  $d(t)$ is the proper distance at epoch $t$. Hence $a(t)>0$. Therefore, we shall consider
  only positive solutions $a(t)$ of Friedmann equations.

%-------------------------------------

  \subsection{Cosmological parameters}

  The Hubble parameter $H(t)$ and the deceleration parameter $q(t)$  are
  defined in Cosmology by
   \begin{equation}\label{DEFHq}
     H(t)= \frac{\dot a(t)}{a(t)}\, ,\quad q(t)= - \frac{\ddot a(t)}{a(t)}\cdot\frac{1}{H(t)^2}
 \end{equation}
  where $a(t)$ is the scale factor. Then obviously we have the
  following identity.
  %Therefore, the equation (\ref{AQL2}) can be written as

  %\begin{equation}\label{AQL3}
  %   \ddot a + \frac{q(t)(H(t)t)^2}{t^2} \hskip 1mm a=0,
  %\end{equation}
  %hence
  \begin{equation}\label{DEFmu}
     \mu(t)= q(t)(H(t)t)^2.
  \end{equation}
  Observe that $\mu(t)$ is a dimensionless parameter.
  We shall assume that $\mu(t)$ is continuous.
  From the physical point of view,                      %taking into account the definition of $\mu(t)$,
  it means that scenarios such as Big Crunch, or
  Big Rip are not included in our analysis. That is, in finite time $t$,
  $a(t)\not = 0$, neither $a(t)$ becomes infinite.
  If  $\mu(t)$ is an $\M$-function then the real constant $\Gamma$ is defined by $\Gamma=\bM(\mu)$.

  Let us remind that the density parameter $\Omega(t)$ and
  the density parameter for the cosmological constant $\Lambda$ are
  defined by
  $$
      \Omega= \Omega(t)= \frac{\rho(t)}{\rho_c}\, , \quad \Omega_{\Lambda}= \Omega_{\Lambda}(t)= \frac{\Lambda}{3H(t)^2}
  $$
  where $\rho_c$ is the critical density.

  %According to   Theorem 3.1, the Friedmann acceleration equation (\ref{AQ}) has two different
  %fundamental solutions  that satisfy  a po\-wer law if and only if the limit
  %\begin{equation}\label{gamma}
  %   \Gamma = \lim_{x\to\infty} x\int_x^{\infty} \frac{\mu(t)}{t^2}dt
  %\end{equation}
  %exists and $\Gamma < \frac{1}{4}$.

  \begin{proposition}
  If the limit $\displaystyle H_\infty = \lim_{t\to\infty} H(t)$ exists, then
  %\begin{equation}%\label{lambdaH}
  $$
   \Gamma=\lim_{t\to\infty}t\left( (H(t) - H_\infty) - \int_t^\infty H(t)^2dt\right)
  $$
  %\end{equation}
  \end{proposition}

  \noindent{\bf Proof.}
  As $\displaystyle\mu(t)= -\frac{\ddot{a}}{a}t^2$ by use of partial integration we have:
  %$$
  %  \int\frac{\mu(t)}{t^2}dt = -\int\frac{\ddot{a}}{a}dt = -\int\frac{d\dot{a}}{a}= -\frac{\dot{a}}a + \int\dot{a}d(1/a)=
  %  -\frac{\dot{a}}a - \int\frac{\dot{a}^2}{a^2}dt
  %$$

  $$
    \int\frac{\mu(t)}{t^2}dt =  -\frac{\dot{a}}a - \int\frac{\dot{a}^2}{a^2}dt= -H(t) - \int H(t)^2dt
  $$
  and the statement follows as
  $\displaystyle \Gamma= \lim_{x\to\infty} x\int_x^{\infty}\hskip -1mm \frac{\mu(t)}{t^2}dt.$
  \qed
  \medskip

  Therefore, if the limit (\ref{gamma}) exists then $\Gamma$ depends solely on the behavior of the Hubble parameter $H(t)$
  at $\infty$.

  The next theorem describes the main property of the scale factor $a(t)$ for the non-oscillatory universe.
  Namely, it gives the necessary and sufficient condition for $a(t)$ to satisfy the generalized power law.

  \begin{theorem}\label{maintheorem} {\rm (Generalized power law for the scale factor $a(t)$)}
  Let $a(t)$ be the scale factor, a solution of Friedmann equations, and $\alpha\in \mathbb{R}$. Then

  \begin{enumerate}
  \item
  If $\mu\in\M_\Gamma$   and $\Gamma < 1/4$ then there is $L\in\N$ so that $a(t)= t^\alpha L(t)$,
  where $\alpha$ is a root of the polynomial $x^2 - x + \Gamma$.

  \item
  If there is $L\in\N$ so that $a(t)= t^\alpha L(t)$ then $\mu\in\M_\Gamma$, $\alpha^2 - \alpha + \Gamma =0$ and $\Gamma \leq 1/4$.
  \end{enumerate}
  \end{theorem}
  %\vskip 1mm

  \noindent{\bf Proof}\quad 1. This assertion follows immediately from Proposition \ref{MaricT3}.

  \noindent 2.\noindent\quad The next proof follows the ideas presented in [\cite{12}, Section 1.4].
  So, suppose $a(t)= t^\alpha L(t)$, $L\in\N$. By Representation theorem \ref{KFT1} there is $\varepsilon\in\Z$ so that
  $\displaystyle\dot L= \frac{\varepsilon}{t}L$,  hence
  \begin{equation}\label{formula1}
     t\frac{\dot a(t)}{a(t)}= \varepsilon(t) + \alpha,\quad \left(t\frac{\dot a(t)}{a(t)}\right)^2 = \eta(t) + \alpha^2,\, \eta\in\Z.
  \end{equation}
  Using $\displaystyle \frac{\ddot a}{a}= - \frac{\mu}{t^2}$ and by integration of the identity
  $\displaystyle
     \frac{\ddot a}{a}=  \left(\frac{\dot a}{a}\right)' + \left(\frac{\dot a}{a}\right)^2
  $
   we obtain after multiplying by $x$
  $$
      -x\frac{\dot a(x)}{a(x)} + x\int_x^\infty \left(t\frac{\dot a(t)}{a(t)}\right)^2t^{-2}dt + x\int_x^\infty \frac{\mu(t)}{t^2}dt= 0.
  $$
  By (\ref{formula1}), the last identity and applying $x\to \infty$, we infer $\alpha^2 -\alpha + \Gamma= 0$. Since
  $\alpha$ is a real number, for the discriminant $\Delta= 1-4\Gamma$ of the polynomial $x^2 -x + \Gamma$ must be $\Delta\geq 0$, i.e.
  $\Gamma\leq 1/4$. \qed 
  %\hfill $\square$
  \vskip 1mm

  \noindent{\bf Remark}\, Under certain conditions Theorem \ref{maintheorem}.1 also holds for $\Gamma = 1/4$, i.e. $\alpha=1/2$.
  This case will be discussed in  Section \ref{adjacent}.
  \vskip 1mm

  \begin{theorem}\label{Lemma1}
   Assume $\mu\in\M_\Gamma$  where $\Gamma < 1/4$.
   Let $a(t)= t^\alpha L(t)$ be the corresponding scale factor, where $\alpha\not=0$ and $L\in\N$, with $L$ having the
   $\varepsilon$-representation as in Theorem {\rm \ref{KFT1}}, where $h(t)$ is a positive constant. Then
   \begin{enumerate}
     \item The possible values of the   curvature index $k$ are $0$ and $-1$, i.e. the Friedmann model of the universe is non-oscillatory.
     \item The Hubble parameter $H(t)$ has the following representation
          \begin{equation}\label{Hubble2}
             H(t)= \frac{\alpha}{t} + \frac{\varepsilon}{t}.
          \end{equation}
     \item The deceleration parameter $q(t)$ has the following representations
            \begin{equation}\label{DecelerationParameter1}
                q(t)= \frac{\mu(t)}{\alpha^2}(1+\eta),%\quad \eta\in\Z.
            \end{equation}
            \begin{equation}\label{DecelerationParameter2}
                q(t)= \frac{1-\alpha}{\alpha} -\frac{t\dot \varepsilon}{\alpha^2}(1+\eta) + \tau, \quad \eta,\tau\in \Z.
            \end{equation}
            %for some $\eta,\tau\in\Z$
     %\item  $q\in \M_{(1-\alpha)/\alpha}$ i.e.
            \begin{equation}\label{DecelerationParameter3}
                 q(t)= \frac{1-\alpha}{\alpha} - t\dot \xi + \zeta, \quad \xi, \zeta \in \Z
     %           \bM(q) = \frac{1-\alpha}{\alpha}
            \end{equation}
   \end{enumerate}
  \end{theorem}

  \noindent{\bf Proof}\quad
  {1}. $\N$-functions belong to so called Zygmund class (Bojani\'c and Karamata, see \cite{2}), hence, since $\alpha\not=0$,
   the scale factor $a(t)$ is ultimately monotonous function. Thus, the universe is non-oscilatory,
   hence $k=0$ or $k=-1$.
   \vskip 1mm

  \noindent{2}.\quad The  representation (\ref{Hubble2})   follows from Proposition \ref{MaricT3}
  \vskip 1mm

  \noindent{3}.\quad
  $\displaystyle
     q(t)= -\frac{\ddot a}{a}\cdot \frac{1}{H^2} = \frac{\mu}{t^2}\cdot \frac{1}{\left(\alpha/t + \varepsilon/t\right)^2}=
     \frac{\mu}{\alpha^2}(1+\varepsilon/\alpha)^{-2}= \frac{\mu(t)}{\alpha^2}(1+\eta(t))
  $

  for some  $\eta\in\Z$, i.e. (\ref{DecelerationParameter1}) holds.
  Further, by $\varepsilon$-representation for $\mu(t)$,  Theorem \ref{RTM}, there is $\delta\in \Z$
  so that
  $
     q(t) = (\Gamma - t\dot\varepsilon + \delta)(1+\eta)/\alpha^2.
  $
  As $\Gamma= \alpha(1-\alpha)$ we obtain (\ref{DecelerationParameter2})
   taking $\tau = \Gamma\eta/\alpha^2 + \delta(1+\eta)/\alpha^2$.

  Finally we show that  $q\in \M_{(1-\alpha)/\alpha}$.  According to Theorem \ref{RTM}
  this will prove the representation  (\ref{DecelerationParameter3}). So, we have
  $$
     q(t)= \frac{\mu}{\alpha^2}(1+\varepsilon/\alpha)^{-2}=
     \frac{\Gamma - t\dot\varepsilon + \delta}{\alpha^2} (1+\varepsilon/\alpha)^{-2},\quad
  $$
  hence for $v(t)= (1+\varepsilon(t)/\alpha)^2$ we have
  $$
     \bM(q)= \frac{\Gamma}{\alpha^2}\bM\left(1/v\right) -
     \frac{1}{\alpha^2}\bM\left(t\dot\varepsilon/v\right) +
     \frac{1}{\alpha^2}\bM\left(\delta/v\right).
  $$
  Further,
  $\displaystyle
     \bM\left(1/v\right)= 1
  $
  since
  $\displaystyle
     x\hskip -1.5mm\int_x^\infty \frac{1}{(1+\varepsilon/\alpha)^2}\cdot \frac{dt}{t^2} \to 1\quad {\rm as}\,\, x\to\infty.
  $
  \vskip 1mm

  $\displaystyle
     \bM\left(t\dot\varepsilon/v\right)= 0
  $
  since
  \vskip 1mm

  $\displaystyle
     x\hskip -2mm\int_x^\infty \frac{t\dot\varepsilon}{v}\cdot \frac{dt}{t^2} =
     -x\alpha\hskip -1.5mm\int_x^\infty \frac{1}{t}d\frac{1}{1+\varepsilon/\alpha}=
     \frac{\alpha}{1+\varepsilon(x)/\alpha} - x\alpha\hskip -1.5mm\int_x^\infty \frac{1}{1+\varepsilon/\alpha}\cdot\frac{1}{t^2}dt=
  $

  $\displaystyle
     \frac{\alpha}{1+\varepsilon(x)/\alpha} - \alpha + o(1) \to 0
  $
  as $x\to \infty$.
  \vskip 1mm

  $\displaystyle
     \bM\left(\delta/v\right)= 0
  $
  since
  $\displaystyle
     %x\hskip -1.5mm\int_x^\infty \frac{\delta}{(1+\varepsilon/\alpha)^2}\cdot \frac{dt}{t^2}\to 0\quad {\rm as}\,\, x\to\infty,
     \delta(t)(1+\varepsilon(t)/\alpha)^{-2}
  $
  is a $\Z$-function.
  \vskip 1mm

  Therefore $\bM(q) = \Gamma/\alpha^2 = (1-\alpha)/\alpha$. \qed
   %\hfill $\square$
  %\vskip 2mm

  Now we introduce a new constant $w$ related to the scale factor $a(t)$ which satisfy the
  generalized power law. It will appear that $w$ is in fact the equation of state parameter.
  So assume $a(t) = t^\alpha L(t)$,  $L\in\N$ and $\alpha\not=0$.
  We define $w$ by
  \begin{equation}\label{PSW}
    w\equiv w_\alpha= \frac{2}{3\alpha} -1\quad {\rm ( equation\,\, of\,\, state\,\, parameter)}.
  \end{equation}

  \noindent
  Note that $w\not= -1$.
  As $\Gamma= \alpha(1-\alpha)$, we   have the following statement:

  \begin{proposition}\label{PPSW}
     \begin{enumerate}

     \item  $\displaystyle \Gamma = \frac{2}{9} \cdot \frac{1+3w}{(1+w)^2}.$

     \item  $\displaystyle w= \frac{1-3\Gamma + \sigma_\alpha \sqrt{1-4\Gamma}}{3\Gamma}$, where $\sigma_\alpha\in\{1,-1\}$.
     \hfill $\square$

     \end{enumerate}
  \end{proposition}
  %\vskip 2mm

  The sign $\sigma_\alpha$ is determined as follows. Suppose $\Gamma\not=1/4$. Then the polynomial $x^2-x+\Gamma$ has
  two different roots $\alpha$, $\beta$. As $\alpha + \beta=1$, we see that $\alpha>\beta$ if and only if $\alpha>1/2$.
  Since $w$ in decreasing in $\alpha$ we have:
  \vskip 2mm

  \noindent{\it Case} $\alpha>1/2$. Then: if $1/4>\Gamma >0$ then $\sigma_\alpha= -1$; if $\Gamma<0$ then $\sigma_\alpha= +1$.

  \noindent{\it Case} $\alpha<1/2$. Then: if $1/4>\Gamma >0$ then $\sigma_\alpha= +1$; if $\Gamma<0$ then $\sigma_\alpha= -1$.
  \vskip 2mm

  According to Theorem \ref{Lemma1} we have also the following statement

  \begin{theorem}\label{Lemma2}
     Under the assumptions of Theorem {\rm \ref{Lemma1}} there are the following relations
     \begin{equation}\label{wparamaters}
     \begin{array}{rlrlrl}\hskip -8mm
       \alpha&= \displaystyle\frac{2}{3(1+w)},                   &a(t)&= \displaystyle a_0L(t)t^{\frac{2}{3(1+w)}}     \\[10pt]
       \hskip -3mm
       %\rho  &\sim \displaystyle\frac{1}{6\pi G(1+w)^2t^2},    &p&\sim wc^2\rho                     \\[10pt]
      H(t)  &\sim \displaystyle\frac{2}{3(1+w)t},    &\bM(q)   &= \displaystyle\frac{1+3w}{2}  \hfill\square
   \end{array}
   \end{equation}

  \end{theorem}
  \vskip 2mm

  For determination of energy density $\rho(t)$ and pressure $p(t)$
  more information on the geometry of the universe are needed. We proceed to study cosmological
  parameters of the universe with the specific curvature index $k$.

 %__________________________________________________________________

  \subsection{Asymptotic solution for universe with curvature index $k=0$}

  In this subsection we shall discuss cosmological parameters for spatially flat universe.
  Hence $k=0$ where $k$ is the curvature index. We also assume
  that the scale factor $a(t)$ satisfies the generalized power law. This allows us to estimate at infinity
  parameters $\rho=\rho(t)$ and $p=p(t)$. The symbol $w$ denotes the equation of state parameter as defined in
  the previous subsection.

  \begin{theorem}\label{Lemma3}
   Assume $\mu\in\M_\Gamma$  where $\Gamma < 1/4$.
   Let $a(t)= t^\alpha L(t)$ be the corresponding scale factor, where $\alpha\not=0$ and $L\in\N$, with $L$ having the
   $\varepsilon$-representation as in Theorem {\rm \ref{KFT1}}. Then

   {\rm 1}.\, $\displaystyle\rho= \frac{1}{6\pi G(1+w)^2t^2} + \frac{\eta}{t^2}$, \quad $\eta\in\Z$.\quad
   {\rm 2}.\, $\displaystyle \bM\left(\frac{p}{\rho c^2}\right) = w$.

  \end{theorem}

  \noindent{\bf Proof}\, 1. As $k=0$, the Friedmann equation (\ref{Friedman}) becomes
  $\displaystyle H^2 =  8\pi G\rho/3$. As $\displaystyle H(t) = \alpha/t + \varepsilon/t$ and
  $\displaystyle w= \frac{2}{3\alpha} - 1$, the
  statement follows if $\displaystyle \eta= \frac{3(2\varepsilon + \varepsilon^2)}{8\pi G}$.
  \vskip 2mm

  \noindent 2.\, By (\ref{mu}), Theorem \ref{RTM}  and the above representation of $\rho$, we have
  $$
    \frac{p}{\rho c^2} = \frac{2\mu}{3\alpha^2(1 + \varepsilon/\alpha)^2} - \frac{1}{3}, \quad \mu= \Gamma - t\dot\varepsilon + \eta,\, \eta\in \Z.
  $$
  Let us take $v(t)= (1 + \varepsilon(t)/\alpha)^2$. Then
  $$
     \bM\!\left(p/\rho c^2\right) = \frac{2\Gamma}{3\alpha^2}\bM\!\left(1/v \right) -
                              \frac{2}{3\alpha^2} \bM\!\left(t\dot\varepsilon/v \right) +
                              \bM\!\left(\eta/v\right) - \frac{1}{3}.
  $$
  As in the proof of Theorem \ref{Lemma1}.3, we have
  $\bM(1/v)=1$, $\bM\!\left(t\dot\varepsilon/v \right)=0$ and  $\bM\!\left(\eta/v\right)=0$.
  Hence, $\bM\!\left(p/\rho c^2\right) = 2\Gamma/3\alpha^2 - 1/3 = 2/3\alpha -1/3 = w$. \qed
  %\hfill $\square$

  \vskip 2mm

 \begin{corollary}\label{wkappa}
         Under assumptions of Theorem {\rm \ref{Lemma3}} there are $\xi,\zeta\in \Z$ so that
         $p= \hat w \rho c^2$, where $\hat w(t)= w -t\dot\xi + \zeta$.
  \end{corollary}

  Hence, the assumption that the scale factor $a(t)$ satisfies the generalized power law implies a
  certain form of equation of state, $p= \hat w \rho c^2$. If $\displaystyle\Gamma=\lim_{t\to \infty} \mu(t)$ exists,
  then $\bM(\mu)=\Gamma$ and by the proof of Theorem \ref{Lemma3} it follows $\hat w = w$, i.e. the classical
  form of the equation of state is valid. In the next subsection we shall see that the assumption of the
  existence of $\displaystyle\Gamma=\lim_{t\to \infty} \mu(t)$ leads to the classical formulas for
  cosmological parameters.

%=======================================================================================

\subsection{Solution for $\mu(t)$ constant at $\infty$}

  In this section we shall discuss conditions under which the parameter $\mu(t)$ introduced by
  (\ref{DEFmu}) is  constant at $\infty$ and how this relates to the solutions
  of the Friedmann equations.
  Therefore  we assume $\displaystyle\lim_{t\to\infty}\mu(t)=\Gamma$. Hence,
  by Proposition \ref{PropositionM}, $\bM(\mu)=\Gamma$. So, all up to now derived properties of
  cosmological parameters related to the scale factor $a(t)$ which satisfies the generalized power law are
  valid. By Theorem  \ref{maintheorem} this will be the case if $\Gamma<1/4$ and under under
  additional assumptions if $\Gamma=1/4$. In this subsection we shall assume $\Gamma<1/4$.
% Observe that if $\mu(t)$ is constant, having the value $\Gamma$,
% then the acceleration equation reduces to the Euler equation
% $ \ddot a + \Gamma a/t^2=0$.
  %If not otherwise stated, we assume for the cosmological constant $\Lambda=0$.
 % We shall consider three cases:
 % $\Gamma<\frac{1}{4}$, $\Gamma=\frac{1}{4}$ and $\Gamma>\frac{1}{4}$.
  \medskip

%__________________________________________________________________

   {\bf Case $k=0$, spatially flat universe}
   \medskip

   By Corollary \ref{EpsC} $\displaystyle\lim_{t\to\infty}\mu(t)=\Gamma$ if and only if $\displaystyle\lim_{t\to\infty}t\dot\varepsilon=0$
   in $\varepsilon$-representation of $a(t)$ described by Theorem \ref{KFT1}. Hence, according to  Theorems
   \ref{Lemma1}, \ref{Lemma2} and \ref{Lemma3} we immediately obtain:
   \begin{equation}\label{paramaAll}
     \begin{array}{rlrlrl}\hskip -8mm
       \alpha&= \displaystyle\frac{2}{3(1+w)},                   &a(t)&= \displaystyle a_0L_\alpha(t)t^{\frac{2}{3(1+w)}}     \\[10pt]
       \hskip -3mm
       \rho(t)  &\sim \displaystyle\frac{1}{6\pi G(1+w)^2t^2},    &p(t)&\sim wc^2\rho                     \\[10pt]
      H(t)  &\sim \displaystyle\frac{2}{3(1+w)t},    &q(t)   &\sim \displaystyle\frac{1+3w}{2}
   \end{array}
   \end{equation}
   and the equations (\ref{AQ}), (\ref{fluid}) and (\ref{Friedman}) are satisfied.

   First we suppose that $\alpha$ is greater of the roots of the polynomial $x^2 - x + \Gamma$, hence $\alpha > 1/2$.
   Then by (\ref{paramaAll}) immediately follows $\displaystyle \frac{1}{3}>w>-1$, hence
   the set of admissible values of $w$ is the interval
   \begin{equation}\label{IalphaB}
     I_\alpha= \left(-1,1/3\right)
   \end{equation}

   The value $w= -1$ yields singularity; for such $w$ there is no corresponding $\alpha$ neither $\Gamma$. If $p= -\rho c^2$ is
   anyway assumed, then by fluid equation we have $\dot{\rho}= 0$, i.e $\rho$ is constant. This case
   corresponds to the cosmological constant, so $\rho= \rho_\Lambda = \frac{\Lambda}{8\pi G}$.
   The constant $\Lambda$ has a negative effective pressure, and as the universe expands, work is
   done on the cosmological constant fluid. Hence energy density remains constant in spite of the
   fact that universe expands.

   If $w=1/3$ then $\alpha= \beta= 1/2$, $\Gamma= 1/4$ and in this case (\ref{GeneralSolution}) is not the general solution
   for Friedmann equations.
   This case will be discussed later.

   If $w= -1/3$, then $\alpha= 1$, $\Gamma= 0$ and the acceleration equation reduces
   to $\ddot{a}\sim 0$. If $\rho$ is computed using the acceleration equation,
   assuming the asymptotic value for $a(t)$ in (\ref{paramaAll}), then  the following asymptotic formula for $\rho$ holds,
   except for $w= -1/3$,
   $$
       \rho\sim \frac{3\Gamma}{4\pi G(1+3w)}\cdot\frac{1}{t^2}
   $$
   Hence, $w= -1/3$ is a kind of singularity, but not the proper one, as it can be replaced by the second
   formula for $\rho$ in (\ref{paramaAll}).

   Let us take into account the physical constraints on the parameters occurring in our calculations.
   For example, the universe is decelerating if and only if
   $q>0$ hence, by (\ref{DecelerationParameter2}), this is equivalent to $\alpha<1$. On the other hand,
   $\alpha<1$ is equivalent to $\displaystyle w>-\frac{1}{3}$ by definition (\ref{PSW}) of $w$.
   Therefore,
   \begin{equation}\label{EQ152b}
     \frac{1}{2}<\alpha<1  \quad \mbox{if and only if} \quad -\!\frac{1}{3}<w<\frac{1}{3}
   \end{equation}
   and the universe decelerates in all cases.
   From $p\sim wc^2\rho$ we see that the pressure $p>0$ if and only if $w>0$.
   Hence, using (\ref{PSW}), we see that $p>0$ if and only if $\alpha<\frac{2}{3}$
   and the interval for $\alpha$ in (\ref{EQ152b}) reduces to $\frac{1}{2}<\alpha<\frac{2}{3}$.

   Let us consider the second fundamental solution  in (\ref{GeneralSolution}) of the acceleration equation
   with the index $\beta\equiv\alpha_2<1/2$.
   First we   introduce the constant $w_\beta$ by
   \begin{equation}\label{constwbB}
         w_\beta = \frac{2-3\beta}{3\beta}
   \end{equation}
   Then  as $\alpha + \beta= 1$ and the following symmetric identity holds:
   \begin{equation}\label{symmetricb}
       w_\alpha + w_\beta+ 3w_\alpha w_\beta = 1
   \end{equation}
   Let ($\mbox{\ref{paramaAll}}_\beta$) be the set of parameters obtained from (\ref{paramaAll})
   by replacing $w(=w_\alpha)$ by $w_\beta$. Using (\ref{symmetricb}) one can show that $b(t)$ satisfies
   all three equations (\ref{AQ}), (\ref{fluid}) and (\ref{Friedman}). Now having $\beta<\frac{1}{2}$
   we can extend the interval for $w$ in (\ref{EQ152b}). As $\beta<\frac{1}{2}$ we find from (\ref{constwbB})
   that $w\equiv w_\beta>\frac{1}{3}$ or $w\equiv w_\beta<-1$.
   Therefore, by  ($\mbox{\ref{paramaAll}}_\beta$)   the set of admissible values for $w=w_\beta$ is the set
   \begin{equation}\label{IbetaB}
     I_\beta= \left(-\infty,-1\right)\cup \left(\frac{1}{3},+\infty\right)
   \end{equation}
   Putting together (\ref{IalphaB}) and (\ref{IbetaB}) we see that the set of all
   admissible values for $w$ corresponding to all possible solutions of the equations (\ref{AQ}), (\ref{fluid}) and (\ref{Friedman})
   is the set
   \begin{equation}\label{IallB}
     I=  \mathbb{R}\backslash \left\{-1, \frac{1}{3} \right\},\quad \mbox{ $\mathbb{R}$ is the set of real numbers.}
   \end{equation}
   The physical constraints on physical parameters for $\beta<\frac{1}{2}$ give the narrower bound
   for $w_\beta$.
   By ($\mbox{\ref{paramaAll}}_\beta$) the adiabatic sound speed
   $v_s=\left( \frac{\partial p}{\partial \rho}\right)^{1/2}= w_{\beta}c^2$,
   hence $w_\beta<1$ and
   so $\beta>\frac{1}{3}$. Therefore, including also our previous discussion on physical constraints
   for $\alpha>\frac{1}{2}$, we have the following bounds for $\alpha$ and $w$:
   \begin{equation}\label{physB}
       \frac{1}{3}<\alpha<\frac{2}{3}\quad \mbox{and}\quad 0<w<1 \quad (\mbox{Zel'dovich interval}).
   \end{equation}
   %\smallskip

% ----------------------------------

      {\bf Case $k=-1$, spatially open universe}
   \medskip

    We shall only briefly discuss this case. We remind that if we assume $\displaystyle\Gamma = \lim_{t\to\infty} \mu(t)$ exists
    and $\Gamma<1/4$, then $t\dot\varepsilon(t)\to 0$ as $t\to\infty$.

    \begin{lemma}\label{LemaPom}\quad
        Let $a(t)= t^\alpha L(t)$ be a solution of Friedmann equations,

        \noindent
        $L\in\N$, with $\varepsilon$-representation {\rm (\ref{RVrepresentation})}
        and $k=-1$ the spatial index. Then
        \begin{equation}\label{ident}
           \frac{p}{\rho c^2}=
           \frac{2}{3}\cdot \frac{\alpha + \varepsilon+ kc^2a^{-2}t^2 -t\dot\varepsilon}{(\alpha + \varepsilon)^2 + kc^2a^{-2}t^2} -1
        \end{equation}
    \end{lemma}
    \vskip 2mm

    \noindent{\bf Proof}\, Taking the logarithmic derivative of $\rho$ using (\ref{Friedman}) we obtain
    $$
    \frac{\dot\rho}{\rho}= 2\cdot\frac{H\dot H - kc^2\dot a a^{-3}}{H^2 + kc^2a^{-2}}
    $$
    Using $\dot a= Ha$ and by (\ref{Hubble2}), $\dot H = \dot\varepsilon/t - (\alpha + \varepsilon)/t^2$, we get
    \begin{equation}\label{OpenW}
           \frac{\dot\rho}{\rho}= -2H\cdot \frac{\alpha + \varepsilon+ kc^2a^{-2}t^2 -t\dot\varepsilon}{(\alpha + \varepsilon)^2 + kc^2a^{-2}t^2}
    \end{equation}
    By fluid equation (\ref{fluid}) we have $\displaystyle p/\rho c^2= -\frac{1}{3H}\cdot \frac{\dot\rho}{\rho} -1$,
    hence (\ref{ident}) follows. \qed
    %\hfill $\square$
    \vskip 2mm

    \begin{corollary}\label{LemaPomC}
        Suppose $k= -1$ and $\alpha<1$. Then $p/\rho c^2 \to -1/3$ as $t\to\infty$.
    \end{corollary}
    \vskip 2mm

    \noindent{\bf Proof}\, As $\alpha<1$ and $L$ is slowly varying, we have $a^{-2}t^2= L(t)t^{2(1-\alpha)}\to \infty$
    as $t\to\infty$. Since also $\varepsilon, t\dot\varepsilon\to 0$ as $t\to\infty$, by (\ref{ident}) the assertion follows. \qed
    %\hfill $\square$
    \vskip 2mm

    \begin{theorem}\,
        Suppose $\displaystyle\Gamma = \lim_{t\to\infty} \mu(t)$ exists, $\Gamma<1/4$, and $a(t)= t^\alpha L(t)$, $\alpha\not= 0$, is
        a corresponding scale factor for an open universe {\rm (}i.e. $k=-1${\rm )}. Then $\alpha=1$ and $w=-1/3$.
    \end{theorem}
    \vskip 2mm

    \noindent{\bf Proof}\, Suppose $\alpha=1$. Then it easy to see that in this case, by (\ref{ident}),
    $$
          \frac{p}{\rho c^2}=
          \frac{2}{3}\cdot \frac{1 + \varepsilon+ kc^2L^{-2} -t\dot\varepsilon}{(1 + \varepsilon)^2 + kc^2L^{-2}} -1\to -\frac{1}{3},\quad
          {\rm as}\,\, t\to\infty.
    $$
    Assume $\alpha<1$. Then, by Corollary \ref{LemaPomC}, we have $p/\rho c^2\to -1/3$, as $t\to\infty$.
    Further, as $a^{-2}t^2\to \infty$ for $t\to\infty$, by (\ref{OpenW}) it follows $\dot\rho/\rho\to -2H$, as $t\to \infty$.
    Then by (\ref{Hubble2}) it follows $\rho(t) = L_1t^{-2\alpha}$ for some $L_1\in\N$.
    By (\ref{Hubble2}), Friedman equation (\ref{Friedman}),  some constants $c_1,c_2$ and $L_2\in \N$, $L_2= L^{-2}$, it follows
    \begin{equation}\label{ident3}
       (\alpha + \varepsilon)^2 = (c_1L_1 + c_2L_2)t^{2(1-\alpha)}.
    \end{equation}
    Since $2(1-\alpha)>0$ and $L_1,L_2$ are slowly varying, it follows
    $$
       L_1t^{2(1-\alpha)}, L_2t^{2(1-\alpha)}\to\infty\quad {\rm as}\,\, t\to\infty,
    $$
    contradicting the identity (\ref{ident3}), as $(\alpha + \varepsilon)^2\to \alpha^2$ for $t\to\infty$.
    Thus, $\alpha\geq 1$.

    Suppose $\alpha>1$ and $t\to\infty$. Then  for the second fundamental solution
    $u_\beta$ we would have $\beta<1$ as $\alpha+\beta=1$. But this solution is impossible as
    it was just proved  in the previous case $\alpha<1$.

    Therefore $\alpha=1$. \qed
    %\hfill $\square$
    \vskip 2mm

   Hence, if the power law is assumed for the scale factor $a(t)$ for an open universe, then
   $a(t)\sim a_0t$ as $t\to\infty$. Also, $\Gamma= 0$ since $\Gamma= \alpha(1-\alpha)$.
   Our analysis in this subsection leads to the following conclusions.
    \smallskip

  \noindent{$\bf 1^\circ$}\,{\bf The values of the equation of state parameter $w$}.
  Let us discuss the values of the parameter $w$ excluded by (\ref{IallB}). In the following we shall use the relation
  (\ref{symmetricb}). We see that  $w=-1$ leads to the singularities in (\ref{paramaAll}).
  Also,  $w_\alpha=-1$ if and only if $w_\beta=-1$ and in this case there is no corresponding $\Gamma$.
  This case corresponds to cosmic inflation.
  The value $w=-\frac{1}{3}$ yields a kind of  singularity in (\ref{paramaAll}), while the relation
  (\ref{symmetricb}) is inconsistent. In this case $w_\beta$ does not exists and the symmetry between $w_\alpha$ and
  $w_\beta$ is broken.
  Also $\Gamma=0$ and the corresponding $w=-\frac{1}{3}$ appears in the solution for the open universe.
  If $w=\frac{1}{3}$, then $\Gamma=\frac{1}{4}$, $w_\alpha=w_\beta$ and $\alpha=\beta$. This case will be analyzed
  in the next subsection. Finally, let us consider the cases $w=0,1$, the values that appear as limits in (\ref{physB}).
  If $w=0$, then $\alpha=\frac{2}{3}$ and from the definition of $w$ we see that $p=0$ (the matter dominated universe).
  If $w=1$, then $\alpha=\frac{1}{3}$ and this value of $w$ corresponds to the universe
   with the mixture of dust and radiation. This is possible only if the integration constant $c_2$ in (\ref{GeneralSolution}) of the
   dominant fundamental solution  is equal to 0.
  \smallskip

  \noindent{$\bf 2^\circ$}\,
  By discussion in this subsection and the remarks in $\bf 1^\circ$ we arrive to the following conclusion:
  {\sl For the spatially flat universe the assumption that
       $\mu(t)$ is constant at $\infty$ leads to the classic solution of the Friedmann equation.}
  \smallskip

   \noindent{$\bf 3^\circ$} For the sake of completeness we give a rather
   short derivation of solution for the spatially flat universe assuming the equation of state $p= wc^2\rho$, $w$ is nonsingular.
   Under this assumption
   the acceleration equation is reduced to $\displaystyle\frac{\ddot a}{a}= -\frac{4\pi G}{3}(1+3w)\rho$, while the
   Friedmann equation becomes $\displaystyle\left(\frac{\dot a}{a}\right)^2= \frac{8\pi G}{3}\rho$.
   Dividing the acceleration equation by the Friedmann equation,
   we obtain $\displaystyle\frac{\ddot a}{a}= \lambda\frac{\dot a^2}{a^2}$ where $\displaystyle\lambda= -\frac{1+3w}{2}$.
   Hence  $\displaystyle\frac{d\dot a}{\dot a}= \lambda \frac{da}{a}$, i.e. $\log \dot a= \log(c_0a^\lambda)$ and so
   $\displaystyle\frac{a^{1-\lambda}}{1-\lambda}= c_0t+c_1$, where $c_0,c_1$ are the integration constants.
   Taking $a(0)=0$, we find $\displaystyle a(t)= a_0t^\frac{1}{1-\lambda}= a_0t^\frac{2}{3(1+w)}$ for some
   constant $a_0$.
   \smallskip

   \noindent{$\bf 4^\circ$}\, {\bf Generalized power law principle}.
   %According to the Theorem  \ref{maintheorem}, if the  integral limit
   %(\ref{gamma})
   %\begin{equation} \label{IntMuB}
   %\Gamma=\displaystyle \lim_{x\to\infty} x\hskip -1mm\int_x^{\infty}\hskip -1mm \frac{\mu(t)}{t^2}dt
   %\end{equation}
   %exists then the solutions $a(t)$ of the Friedmann equations satisfy the generalized power law if and only if $\Gamma<\frac{1}{4}$.
   Putting together all results presented up to now, we see that the following are equivalent:
   \smallskip

   \noindent{\bf a.} The integral limit $\Gamma$ in (\ref{gamma}) exists and $\Gamma<\frac{1}{4}$.

   \noindent{\bf b.} The solutions $a(t)$ of the Friedmann equation
   satisfy the generalized power law with index $\alpha\not= 1/2$.

   \noindent{\bf c.} The equation of state holds at infinity as described by
   Corollary \ref{wkappa}, $w\not= -1,  \frac{1}{3}$.
   If $\mu(t)$ is constant at $\infty$  then $p\sim wc^2\rho$, as $t\to\infty$.

   \noindent{$\bf 5^\circ$}\, {\bf Power law principle and cosmological constant $\Lambda$}.
   If $\Lambda\not= 0$ is assumed, all the asymptotic formulas (\ref{paramaAll}) for
   the cosmological parameters are valid, except for $w= -1$.
   This follows from the fact that by the appropriate substitutions the Friedmann
   equations with  the parameter $\Lambda$ transform to their basic form (\ref{Friedman}) -- (\ref{fluid}).

%__________________________________________________________________

   \subsection{Case $\Gamma= \frac{1}{4}$, adjacent case}\label{adjacent}

   In this subsection we shall assume  $\Gamma= \frac{1}{4}$ in the limit (\ref{gamma}).
   Then the polynomial $x^2-x-\frac{1}{4}$  has the double root $\alpha=\frac{1}{2}$.
   In discussion of the acceleration equation (\ref{AQL2}) for this case we shall  use
   the following criterion, see Mari\'c [\cite{12}, p. 37] and Kusano, Mari\'c [\cite{10a}, Theorem 2.2]:
   \smallskip
   \begin{theorem}
   {
   Let $\displaystyle \phi(x)= x\int_x^\infty \frac{\mu(t)}{t^2}dt -\frac{1}{4}$,\,  let   the integral
   \begin{equation}\label{phiB}
     \psi(x)=\int_{x}^\infty \frac{|\phi(t)|}{t}dt,\quad x>x_0>0, \mbox{ converge}
   \end{equation}
   and assume
   \begin{equation}\label{psiB}
       \int_x^\infty \frac{\psi(t)}{t}dt< \infty,\quad x>x_0.
   \end{equation}
   Further, let $L_1, L_2$ denote two normalized slowly varying functions. Then there exist
   two fundamental solutions of the acceleration equation {\rm (\ref{AQL2})}:
   \begin{equation}\label{uv}
     u(t)= t^\frac{1}{2}L_1(t)\quad  v(t)= t^\frac{1}{2}\log(t)L_2(t)
   \end{equation}
   if and only if the condition {\rm (\ref{gamma})} holds {\rm (}for $\Gamma=\frac{1}{4}${\rm )}.
   Also $L_1, L_2$ tend to a constant as $t\to \infty$   and $L_2(t)\sim 1/L_1(t)$.
   }\hfill \qed
   \end{theorem}
   \smallskip

   As $\log(t)L_2(t)$ is also a slowly varying, we see that both fundamental solutions $u(t)$ and
   $v(t)$ satisfy the general form of power law.
   Hence, each solution $a(t)= c_1u(t)+c_2v(t)$ of the acceleration equation is regularly varying
   of index $\frac{1}{2}$.
   By the results in the previous subsection when $\Gamma<\frac{1}{4}$ was assumed, we see,
   if the conditions (\ref{phiB}) and  (\ref{psiB}) are satisfied, that $a(t)$ is regularly varying of index $\frac{1}{2}$
   if and only if $w\sim \frac{1}{3}$ as $t\to \infty$, i.e.  $p\sim \frac{1}{3}c^2 \rho$
   holds asymptotically. This is the second classic cosmological solution.

%__________________________________________________________________

%========================================================================================

 \vskip 5mm

 \subsection{Asymptotic solution for spatially flat universe with matter-dominated evolution}

  We have seen that the constant $\Gamma=\bM(\mu)$ determine the asymptotical behavior at the infinity
  of the  scale factor $a(t)$.
  If the matter-dominated evolution of the universe is assumed, i.e. dominated by some form
  of pressureless material after the certain time moment $t_0$ then it appears that the
  expression $H(t)t$ depends solely on the parameter $\Omega$. In this case we are able to estimate possible
  values of $\Gamma$. We shall discuss also the status of the constant $\Gamma$ and   the related
  asymptotic behavior of $a(t)$ for the spatially flat universe including the cosmological constant
  $\Lambda$. Therefore, in this section we  discuss asymptotic solutions and  Friedmann  equations, the related parameter $\mu(t)$,
  and the constant $\Gamma$  assuming the pressureless spatially flat universe  with
  the cosmological constant $\Lambda$.

\begin{figure}
  \centering
     \includegraphics{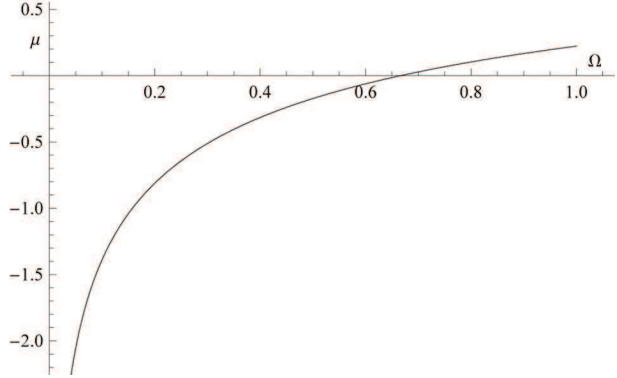}
     \caption{Graph of $\bar\mu(\Omega)$}
  \label{Graph1}
  \end{figure}

  Using the formula for the age of the spatially  flat universe with the cosmological constant $\Lambda$
  Carroll at al. \cite{3}, see also Liddle and Lyth \cite{9} and Narlikar \cite{15},
  the expression $H(t)t$ in this case is  given by
  \begin{equation}\label{flat}
    H(t)t= \frac{2}{3}\cdot\frac{1}{\sqrt{1-\Omega}}\ln\left(\frac{1+\sqrt{1-\Omega}}{\sqrt{\Omega}}\right),
  \end{equation}
  while the deceleration parameter $q(t)$ is given by
  \begin{equation}\label{qt}
     q(t)= \frac{\Omega}{2} - \Omega_{\Lambda}.
  \end{equation}
  In the model of the spatially flat universe we have
  $
     \Omega + \Omega_\Lambda=1
  $
  hence,
  $
    \displaystyle q(t)= \frac{3\Omega}{2} -1.
  $
  Therefore, by (\ref{DEFmu}) and (\ref{flat}) it follows
  \begin{equation}\label{qt2}
     \mu(t)= \frac{2}{9}\cdot\frac{3\Omega-2}{1-\Omega}
             \left(\ln\left(\frac{1+\sqrt{1-\Omega}}{\sqrt{\Omega}}\right)\right)^2.
  \end{equation}
  where $\Omega= \Omega(t)$.
  We see that the parameter $\mu(t)$ in the model for the pressureless spatially flat universe depends solely on $\Omega$.
  The graph of the parameter $\mu(t)$ is presented   in Figure \ref{Graph1}, as a
  function of $\Omega$.

  The limit value
  \begin{equation}\label{EQ24}
    \Omega_{\infty}= \lim_{t\to\infty}\Omega(t)
  \end{equation}
  can be in principle any value in the interval $[0,1]$.
  Let us introduce the parameter $\bar\mu(\Omega)$ by
  the expression on the right hand side of (\ref{qt2}).
  Hence $\mu(t)=\bar\mu(\Omega(t))$ and
  \begin{equation}\label{EQ27}
    \Gamma= \lim_{t\to\infty}\mu(t)= \lim_{\Omega\to\Omega_{\infty}}\bar\mu(\Omega).
  \end{equation}
  We see that $\bar\mu(\Omega)$ is an increasing function in $\Omega$ and that its values lay in the
  interval $[-\infty,\frac{2}{9}]$, as
  $\displaystyle\lim_{\Omega\to 1-0 }\bar\mu(\Omega) = \frac{2}{9}$.
  Hence $\displaystyle \mu(t) < \frac{2}{9}$. Suppose the limit  $\displaystyle\Gamma = \lim_{t\to \infty}\mu(t)$ exists.
  Then $\Gamma\leq\frac{2}{9}<\frac{1}{4}$. Thus, assuming the pressureless spatially flat universe with the cosmological constant,
  all possible values of  $\Gamma$ are less than $2/9$, hence    Theorem \ref{maintheorem} can be applied.
  This analysis leads to the following conclusions for the solutions of Friedmann
  equations for the pressureless  spatially flat  universe with the cosmological constant $\Lambda$.
  \medskip

  \noindent{$\bf 1^\circ$} The scale factor $a(t)$ satisfies the generalized power law.
  More specifically, $a(t)= t^\alpha L(t)$ where $L\in \N$ and $\alpha$ is a root of
  the polynomial $x^2 -x +\Gamma$.
  \medskip

  \noindent{$\bf 2^\circ$} Suppose $\Omega_\infty=1$. By the identity
  $\Omega + \Omega_\Lambda=1$ it follows $\Omega_\Lambda\sim 0$ as $t\to\infty$ and
  $\Gamma=\frac{2}{9}$. Then the equation (\ref{EQ144})   becomes
  $
    x^2-x+\frac{2}{9}= 0
  $
  and it has the solutions $\alpha_1=\frac{1}{3}, \alpha_2=\frac{2}{3}$. According to
  the conclusion $\bf 1^\circ$, $a(t)$ regularly varying of index $\frac{2}{3}$ and by (\ref{flat}) and (\ref{qt}),
  $H(t)t\sim \frac{2}{3}$ and $q\sim \frac{1}{2}$ as $t\to\infty$.
  This result corresponds to the classic solution of the Friedmann equation
  for the  pressureless spatially flat universe with the cosmological constant $\Lambda=0$.
  \smallskip

  \noindent{$\bf 3^\circ$} The formula (\ref{qt2}) for $\bar\mu$ shows that
  the evolution of the expansion scale factor $a(t)$
  depends only on the evolution of the density parameter $\Omega$.
   The nature of this evolution is determined by the constant $\Gamma$ but in all
  instances it satisfies the power law represented by some regularly varying
  function.  The introduction of the cosmological constant only changes the
  index of the regular variation with respect to the model with $\Lambda=0$.
  \medskip

  \begin{figure}
  \centering
     \includegraphics{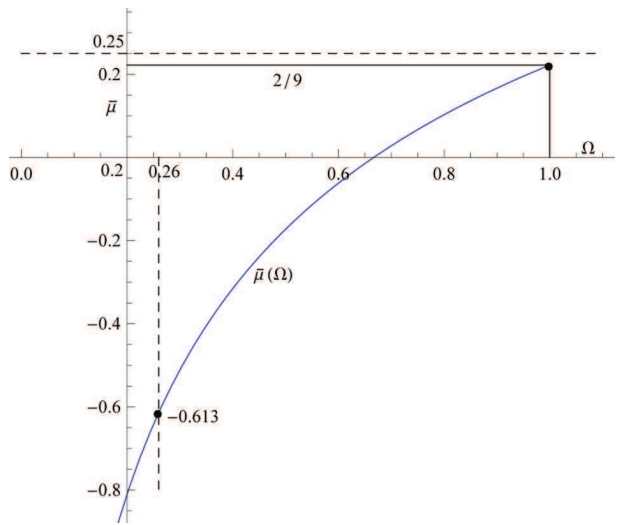}
     \caption{Graph of $\bar\mu(\Omega)$ in the preferred interval }
  \label{Graph2}
  \end{figure}

  \noindent{$\bf 4^\circ$} Let us consider the possible values of  $\Gamma$.
  The value $\Omega_0= 0.3$ (e.g. Liddle \cite{9}) for the present epoch is close to the value preferred by the
  observation. If we assume that the energy density $\rho$ becomes lower as the age of
  the universe becomes older,
  we may suppose that the possible range for the constant $\Omega_{\infty}$ is
  the interval $[0.3,1]$, i.e. $0.3\leq\Omega_\infty \leq 1$.
  The graph of $\bar \mu$ for this interval is presented in Fig. 2. We see that $\Gamma\leq 2/9$.

  \noindent{$\bf 5^\circ$} The solution $a(t)$ is regularly varying of
  some index $\alpha$, i.e. $a(xt)/a(t)\to x^\alpha$ as $t\to\infty$, $x>0$. So for relatively
  large\footnote{
  This notion can be made more precise by use of {$nonstandard\ analysis$}. There we can say
  that $t_0$ is infinite, while $x>0$ is non-infinitesimal finite real number. For development
  and notions of nonstandard analysis one can see Stroyan \cite{19}.
  There are a lot of applications of nonstandard analysis in the theoretical physics, e.g. Albeverio \cite{1}.}
  $t_0$ with respect to $x>0$,  we have $a(xt_0)\sim a(t_0)x^\alpha$ and we may take that the time instances
  $t_0$ and $xt_0$ belong to the same epoch in the evolution of the universe.
  So, taking $t=xt_0$ and eliminating $x$ from the last asymptotic relation,
  we find the asymptotic estimation for $a(t)$ for an epoch with respect to
  the initial value $a(t_0)$:
   \begin{equation}\label{AS25}
    a(t)\sim a(t_0)\left(\frac{t}{t_0}\right)^\alpha.
  \end{equation}
  Also, $a(t)= t^{\alpha}L(t)$ where $L(t)$ is slowly varying. Hence
  $L(t)\sim L(t_0)$ for an epoch, so it is hard to measure $L(t_0)$ and
  the influence of $L(t)$ on $a(t)$. However, the influence of $L(t)$ on
  the large scale might be substantial, particularly if $\alpha \approx 0$,
  as the example (\ref{erdos}) shows.
  \medskip

%__________________________________________________________________

    \subsection{Case $\Gamma> \frac{1}{4}$}

    Assume $\Gamma> \frac{1}{4}$ in the limit (\ref{gamma}). Then the solution
    $a(t)$ of the acceleration equation is oscillatory. This immediately follows from Hille's
    classical  theorem (see Hille \cite{6} and Mari\'c \cite{12}, Theorem 1.8). Therefore,
    for these values of $\Gamma$ the expansion scale factor $a(t)$ of the universe
    does not satisfy the power law and the approach presented in the paper is not
    appropriate for this case. Since $a(t)$ has in this case (infinitely many) zeros, then
    there is $t_0$ such that $\dot a(t_0)=0$. So, from the Friedmann equation (\ref{Friedman})
    it follows that $k>0$, i.e. the universe must be closed. This is obviously true even if
    the Friedmann equation is modified by adding the cosmological constant $\Lambda>0$.

    We see that the constant 1/4 plays an important role as a possible value of $\Gamma$ in the limit (\ref{gamma}).
    This constant provides a sharp "threshold",
    or "cut-off point", at which the oscillation of $a(t)$ takes place.

%__________________________________________________________________
%__________________________________________________________________

   \section{Conclusion}

 It has been shown that a dimensionless constant $\Gamma$
 related to the Friedmann acceleration equation  and the theory of regularly varying functions play the key role in the formulation
 of the power law principle for solutions $a(t)$ of the Friedmann equations.
 The constant $\Gamma$ is defined by
 \begin{equation}\label{Gamma3}
       \Gamma=\displaystyle \lim_{x\to\infty} x\hskip -1mm\int_x^{\infty}\hskip -1mm \frac{\mu(t)}{t^2}dt.
 \end{equation}
 where $\mu(t)= q(t)(H(t)t)^2$ as $t\to \infty$,
 $q(t)$ is the deceleration parameter and $H(t)$ is the Hubble parameter.
 We have shown that the generalized power law principle for the scale factor $a(t)$ holds if and only if the integral limit (\ref{Gamma3})
 exists and $\Gamma<\frac{1}{4}$. The cosmological constants were also discussed under relaxed condition
 $\lim_{t\to\infty}\mu(t)=\Gamma$ which implies (\ref{Gamma3}),
 Under this condition we have shown that the power law principle is equivalent to
 the equation of state $p\sim wc^2p$, $w\not= -1, 1/3$.
 The values of $\Gamma$ determine the asymptotical behavior  of the  scale factor $a(t)$   as time $t$ tends to $\infty$.
 The constant $\Gamma$ also uniquely determines
 other cosmological parameters such as the Hubble parameter and the equation of state parameter $w$.
 Particularly is discussed the pressureless spatially flat universe with non-zero cosmological constant.
 Further, the  value of $\Gamma$ determines the type of the universe; for $\Gamma<\frac{1}{4}$
 the universe is spatially flat or open, while for $\Gamma>\frac{1}{4}$ the  universe is oscillatory.
 The boundary case $\Gamma=\frac{1}{4}$  is also analyzed.
 All solutions we found are in agreement with the results found widely in the literature on standard cosmological model.
 As  power law functions  are the most frequently occurring type of the solutions of the Friedmann equation, the study
 of the constant $\Gamma$ and the related function $\mu(t)$  might be of a particular interest.
\bigskip
%__________________________________________________________________
%__________________________________________________________________

\noindent
{\bf Note}\quad
Mathematical contributions in this paper belong to the first author.
The other coauthors clarified physical aspects of this work.
We are particularly grateful to the anonymous reviewers  for their suggestions and
comments. Due to their remarks and suggestion we did the major  revision of the first
version of the paper, particularly those parts concerning the integral condition (\ref{gamma}).
\bigskip

%{\bf Bibliography}
%\smallskip

\end{document}